# Do we need to revisit the Bohr exciton radius of hot excitons?

*Padmashri V. Patil\* and Shouvik Datta\*\**

Department of Physics, Indian Institute of Science Education and Research-Pune,

1st Floor, Central Tower, Sai Trinity Building, Pashan, Pune – 411021, Maharashtra, India.

Email : *padmashriv.patil@gmail.com, **shouvik@iiserpune.ac.in,

## ABSTRACT

We report collisional broadening of the $E_3$ excitonic resonances in optical absorption spectra of PbS nanocrystallites of widely varying sizes. Significance of the underlying extended band structure of bulk solids to understand the physics of exciton scattering in semiconductor nanocrystallites of this size range is discussed. We propose an empirical notion of *'effective Bohr exciton radius'* as a direct consequence of significant departure from usual static screening limits of coulomb interactions of 'hot' excitons in the region of strong dispersion at energies much above the fundamental band gap. Temperature variation of excitonic resonance reveals how non-phonon energy relaxation processes affects the collisional broadening of $E_3$ resonance over this size range. We argue that ballistic transport of hot excitons can suppress exciton-exciton scattering necessary for photo-induced impact ionization in very small quantum dots. This evidently indicates that impact ionization of hot excitons may be efficient only inside an intermediate nanoscale *'size window'*.

1.   INTRODUCTION.

In solids, attractive coulomb interaction can produce bound states of electron-hole pairs called excitons during optical absorptions. Most spectral studies [1-4] of semiconductor nanoparticles were focused primarily on the lowest excitonic transitions where one can comfortably neglect the presence of strong dispersions in dielectric response. Here, we report collisional broadening of excitonic resonance in lead sulfide (PbS) nanocrystallites in room temperature optical absorption spectra at excitation energies (≥5.9eV) much higher than its bulk band gap (~0.4eV). Statistical description of these highly energetic photogenerated carriers maintains an effective temperature ( $T_{Hot\ Carrier}$ ) higher than the equilibrium temperature ( $T_{Lattice}$ ) of the material, thereby calling these 'hot' excitons. However, femto second time domain studies at such photoexcitation wavelengths (~210nm) are not very commonplace and currently out of scope for this work. As an alternative, we have investigated absorption spectra of these PbS nanocrystallites over a wide range nanocrystallite sizes starting from large particles with bulk like band structure to strongly confined particles of very small sizes with discrete density of states. It is also understood that breakdown of momentum conservation for excitons confined inside these nanoparticles can allow such 'hot' excitonic states to have finite center-of-mass momentum K = $k_e$ + $k_h$ and center-of-mass kinetic energy, where $k_e$ and $k_h$ are momentum of electrons and holes.

Bohr exciton radius of such bound electron-hole pair is defined as $a_B = \frac{4\pi\varepsilon_\infty \hbar^2}{\mu e^2}$, where $\varepsilon_\infty$ is the usual optical frequency dielectric constant of the material, e is electronic charge, $\hbar$ is the reduced Plank's constant, $\mu = \left(\frac{m_e^* m_h^*}{m_e^* + m_h^*}\right)$ is reduced mass of electron-hole bound state, $m_e^*$ and $m_h^*$ are effective mass of electron and holes respectively in M.K.S units. This $a_B$ symbolizes the characteristic length scale to observe quantum effects in nanomaterials. In connection with the observed $E_3$ excitonic resonance in PbS, we explore the following issues – (a) Is the above mentioned classic definition of Bohr exciton radius ($a_B$) precise enough to describe transitions involving 'hot' excitons much above the fundamental band gap? (b) How to quantify the dispersive dielectric response in the coulomb interaction of electron-hole bound pair for hot excitons? (c) What are the implications of a dynamically screened effective Bohr radius for exciton-exciton scattering inside nanoparticles? We will try to look forward to find answers to these questions on hot exciton physics throughout this report.

Moreover, it is well known [5,6] that carrier multiplication (CM) or multiple exciton generation (MEG) are predicted to be efficient for hot excitons confined within quantum structures under high energy photo-excitations. Solar cells using such nanoparticles are expected to exploit MEG/CM by generating more than one electron-hole pairs from single photon to enhance the power conversion efficiency beyond the maximum attainable thermodynamic efficiency [7]. Efficient MEG/CM was subsequently reported by several groups [8,9] in many semiconductor nanoparticles. However, recent reports debated [10-12] precise nature of efficiency of MEG/CM in nanoparticles as compared to bulk. Here, we will discuss the important role of collisional broadening of 'hot' excitons via 'zero phonon coupling' to the 'extended' band structure of PbS over wide range of sizes and measurement temperatures. In this context, we will elaborate how the any empirical re-definition of Bohr exciton radius may affect the mean free path of exciton-exciton collision and the basic understanding of MEG/CM. We will argue in favor of a minimum size cut-off of MEG/CM as a result of significant reduction of collisional broadening in the strong confinement [13] regime $\left( radius < \frac{1}{3} a_B \right)$, where the nanocrystallites can be quasi-ballistic for inverse Auger processes.

Unmistakably, such an effective size window for MEG/CM was not reported so far in earlier studies which were mostly focused on so called strongly confined very small nanoparticles only. We will discuss how the gradual size variations of these nanocrystallites from bulk like weakly confined particles to strongly quantum confined particles can affect the excitonic absorption spectra. Size dependence of not only the energy broadening and amount of blue shift of excitonic resonance but also its overall aging and temperature variation will be presented to support our arguments. This will highlight the crucial roles of nanocrystallite sizes and the importance of Auger like non-phonon energy relaxation mechanisms manifested as collisional broadening in the physics of 'hot' excitons. MEG/CM are direct consequences of hot-exciton physics and therefore, our results on the effects of excitonic collisions on absorbance spectra of PbS nanoparticles, although observed at energies much above the visible range of solar spectrum, may help us in understanding the physics of MEG/CM for photovoltaic applications. Our analyses are further substantiated by somewhat similar temperature dependence and aging behavior in PbSe nanoparticles too. Slight difference between PbSe and PbS originates only from the smaller excitonic binding energy and lesser band gap of PbSe as compared to PbS. This makes excitons in PbSe to experience enhanced coulomb screening effects due to the presence of larger number of free carriers. For comparison, all results on PbSe are given in the section VI of the supplemental materials. Care has certainly been taken to ensure that observed results are not coming from any capping ligands or any other chemical compounds like water, reagents etc used in the synthesis.

## 2. RESULTS AND DISCUSSIONS.

**A) Size determination of nanocrystallites and evidence of excitonic resonance in optical absorbance:** The X-ray diffraction (XRD) spectra (figure 1a) of PbS nanoparticles establishes crystalline nature of cubic structure of PbS. We have used Debye-Scherrer method to estimate the mean nanocrystallite diameters from the XRD spectra and will refer this diameter as the size of these PbS crystallites throughout this report. Figure 1b shows room temperature optical absorption spectrum of freshly prepared uncapped PbS nanocrystallites dispersed in de-ionized water. Further details of synthesis procedures of PbS nanoparticles with and without Thio-glycerol (TG) capping ligands, purification processes, control experiments, low energy $E_0$ spectra of PbS nanoparticles, all studies of crystallite size estimations from XRD and hydrodynamic sizes by dynamic light scattering (DLS) as well as representative results on PbSe etc are given in the supplemental material.

In the optical absorption spectra, we see a strong resonance (~5.9eV≈ $1.1 \times E_3^{Bulk}$) just above $E_3$ critical points [14,15] of bulk PbS ($E_3^{Bulk}$ = 5.3eV) in this weakly [13] confined quasi-bulk PbS nanocrystallites with mean diameter ~24nm. Absorption edges are also observed in this quasi–bulk sample near 3.5eV and 2.1eV which are slightly above the respective $E_2$ (3.14eV) and $E_1$ (1.94eV) critical point transitions of bulk PbS. Similar absorption spectra for the lowest $E_0$ transition given in the supplementary material. Understandably, excitonic resonances around low energy critical points, such as $E_0$, in such small band gap material like PbS can easily be suppressed [14] by strong coulomb screening in presence of large numbers of charge carriers near the band edge. Although, existence of bulk/surface electronic defects can decrease free charge carrier density and thereby reduce the effect of such coulomb screening on excitonic resonances. It should be noted that similar excitonic resonances at energies much higher than the fundamental band gap was also reported [14] earlier in optical reflectivity studies of bulk PbS crystals at low temperatures. Moreover, as discussed above, momentum is not a good quantum number for strongly confined [13] nanoparticles, but the residual presence of bulk like extended band structure on the density of states of PbS nanoparticles over the wide size range studied here is not fully unexpected in quasi bulk nanoparticles. Such lingering effects of bulk band structure on the energy levels of semiconductor nanoparticles were predicted [16] theoretically under the truncated crystal approximation too.

**B) Size variation and blue Shift of $E_3$ Resonance:** In figure 1c, we compare the optical absorption spectra of uncapped and TG capped PbS nanocrystallites (having $a_B \sim 18nm$) of various diameters. Absorbance peak of $E_3$ exciton resonance blue shifts and increases monotonically with decreasing nanocrystallite size, which indicate that oscillator strength for excitonic absorption is still size dependent in PbS nanocrystallites even under weak [13] confinement $(radius \sim 3a_B)$. The above enhancement of optical absorbance with decreasing size clearly says that the contribution of light scattering is not significant. Moreover, observed size dependence and blue shift also rules out the TG ligands or any other floating molecular species other than solid PbS as the cause of this $E_3$ excitonic resonance. We see that $E_3$ spectrum of this weakly confined PbS nanocrystallites having 24nm crystallite diameter is gradually broadened into an absorption edge in ~16nm PbS. Yet, below 16nm, we observe progressive sharpening with decreasing size. Further understanding of this unusual evolution of excitonic broadening with crystallite size will be presented near the end of Section 2D and in the beginning of Section 2E. These spectra are fitted with Voigt line shapes from which we estimate both inhomogeneous width (Gaussian) and homogenous width (Lorentzian) of the $E_3$ resonance. Lorentzian widths are nearly ~ 1.5 times the respective Gaussian widths for particles ≥16nm. In general, magnitude of excitonic transition is inversely proportional to its broadening parameter. Nevertheless, inhomogeneous width of $E_3$ exciton is not monotonically decreasing with size (figure 2a). Nominal use of Heisenberg uncertainty relation $\Delta E \, \Delta t \sim \hbar$ with homogeneous broadening ($\Delta E$) estimates the life time ($\Delta t$) of such states around femto-seconds domain.

Moreover, the magnitude of maximum possible energy broadening width $\Delta E_{max}$ = [$\Delta E(R_{Min})$ - $\Delta E(R_{Max})$] is calculated using $\Delta E = \frac{h^2}{8R^2}\left[\frac{1}{m_e^*}+\frac{1}{m_h^*}\right]-\frac{1.8e^2}{4\pi\varepsilon R}$, where $\Delta E_{max}$ is difference between the energy position for minimum (R=$R_{min}$) and maximum (R=$R_{max}$) nanocrystallite radius, h is the Planck constant and R=D/2 is the radius of the crystallite as estimated by XRD. These $\Delta E_{max}$ values (figure 2a) are well below the estimated line widths of $E_3$ resonances (figure 1c) and have qualitatively different size dependence for all nanocrystallite diameters. Therefore, we can rule out any size distribution effect towards energy broadening of $E_3$ exciton. Surprisingly, excitonic transitions (figure 1c) are much above the $E_3$ critical point of bulk PbS (0.6eV larger than 5.3eV) and also larger than its optical phonon energy [4,6] ~20-27meV. It should also be noted that 5.9 eV energy of $E_3$ excitonic resonance is still < 2 × $E_3^{Bulk}$ at 5.3 eV to observe such hot excitonic effects around the $E_3$ critical point. Although the energy of $E_3$ excitonic resonance at 5.9 eV may be well within the energy continuum of states above the lowest fundamental $E_0$ bandgap of PbS but it is certainly not the case for excitons localized around the much larger $E_3$ gap [14, 16]. Nonetheless, estimates (using $\mu \approx 0.05 m_e^0$, $m_e^0$ is the free electron mass) also shows that this excess energy cannot be explained by confinement induced enhancement of energy $\Delta E_{Confinement} = \frac{h^2}{8R^2}\left[\frac{1}{m_e^*}+\frac{1}{m_h^*}\right]$ alone (figure 2b), except for strongly confined PbS nanocrystallites of mean diameter ≤3nm. So it is tempting to attribute this energy difference to – a) failure of effective mass theory [17-18] at smaller sizes, b) non-parabolicity [19] of the band structure at high energies. However, size dependence of excitonic energy is much slower than $\sim \frac{1}{R^2}$ or $\frac{1}{R}$ type of behavior (figure 3b). Therefore, this extra energy may be credited to excess center-of-mass kinetic energy $\left(\frac{\hbar^2 K^2}{2M}\right)$ of $E_3$ exciton as a result of non-conservation of momentum inside nanoparticles, where M is the translational mass of exciton (M = $m_e^*$ + $m_h^*$) as mentioned in the introduction.

**C) Dispersion of dielectric response at energies much higher than the band gap and the concept of effective Bohr exciton radius:** Generally, high frequency optical dielectric constant[15] of PbS as $\varepsilon = \varepsilon_e + \varepsilon_l = \varepsilon_e(0) \cong \varepsilon_\infty \sim 17$ is used to estimate excitonic parameters, where the subscripts 'e' and 'l' stands for electronic and lattice contributions respectively. Approximating dielectric response with a limiting value like $\varepsilon_e(0) \sim \varepsilon_\infty$ is only valid [15] for $\frac{E_g}{\hbar} \gg \omega \gg \omega_T$, where $E_g$ is the band gap and $\omega_T$ is the frequency of transverse optical phonon. Yet, here we have $\omega_{E_3 \text{ Exciton}} \gg \frac{E_3^{Bulk}}{\hbar} \gg \frac{E_0^{Bulk}}{\hbar}$. So, the standard dielectric response [20] theory based on 'electro-statics' may not be adequate here. Experimentally, this $\varepsilon_\infty$ is estimated from the E→0 limit (called the long wavelength limit) of the measured real part of the dielectric constant assuming that the imaginary part (absorbance) more or less vanishes in that limit. However, in the presence of strong resonant optical absorption as significant contribution from imaginary part of the dielectric response, we have used empirical magnitude of 7.2 for complex $\varepsilon(\omega)$ around 5.9

eV of $E_3$ excitonic resonance. This value of $|\varepsilon(\omega)|$ is estimated from the reported [21] spectroscopic ellipsometry measurements on crystalline bulk PbS. One can certainly measure $|\varepsilon(\omega)|$ of these nanoparticles to get a better estimate, which is currently beyond the scope for wavelengths ≤210 nm. We find that calculated exciton binding energy $|E_{EX}| = \left|-\dfrac{1.8e^2}{4\pi\varepsilon R}\right|$ using $|\varepsilon(\omega)| = 7.2$ are higher than that with $\varepsilon_\infty = 17$ and much larger than optical phonon energy of PbS for all sizes. Reader can see the quantitative details of this variation of excitonic binding energy with size in the Table inside Section III of the supplementary material. This can certainly explain the existence of $E_3$ excitonic resonance in PbS nanocrystallites even at room temperature. Assuming that we can still apply the hydrogenic model for exciton physics at such energies, we approximate an *effective Bohr exciton radius ($a^*_{Ex}$)=7.6 nm* assuming $|\varepsilon(\omega)| = 7.2$ at E~5.9eV, $\mu \approx 0.05 m_e^0$. Here we assume that an empirical dielectric response in terms of $\varepsilon(\omega)$ can be defined even for the smallest nanocrystallites without any additional size dependence and surface polarization effects. This effective Bohr exciton radius ($a^*_{Ex}$)=7.6 nm also shows that quasi-bulk like uncapped particles (figure 1b) with crystallite radius (diameter) of 12nm (24nm) can still be treated in the weak [13] confinement regime $\left(radius \sim 3a_B\right)$. This value [21] of $|\varepsilon(\omega)|$ is much larger around $E_2$, $E_1$, $E_0$ critical points of PbS. Therefore, binding energy of any excitons around $E_2$, $E_1$, $E_0$ critical points can be much smaller than the binding energy hot excitonic resonance at $E_3$ critical point. This certainly explains [14] the not so significant presence of excitonic resonance at low energy critical points due to effectively large coulomb screening effects as also explained in Section 2A.

**D) Absence of temperature variation of excitonic resonance even in weakly confined particles and the significance of non-phonon energy relaxations.** Strong coupling of excitons to optical phonons is usually neglected in lead salts because of quasi symmetrical nature of electron and hole bands near its fundamental band gap ($E_0$). However, electronic band structure near $E_3$ critical point in PbS is definitely not symmetric [14,22] ($m_e^* \neq m_h^*$ ~$E_3$). So, we rather expect the broadening of $E_3$ exciton to be dominated by Fröhlich type strong polar exciton-phonon [20] interactions. Despite this, $E_3$ resonance still survives at room temperature and, most importantly, shows no temperature dependent shift of excitonic peak (figure 4) even for weakly confined (~24nm) particles. Therefore, it seems likely that *'non-phonon'* energy relaxation mechanisms [23] like inverse Auger processes may be dominating over phonons for this highly energetic $E_3$ exciton. Further evidences of this assertion come from temperature dependent evolution of excitonic spectra (figure 4a) of bigger nanoparticles, which can be ascribed to Auger cooling effects. This assertion is further vindicated by total absence of such temperature dependence of absorption spectra in smaller nanoparticles (figure 4c) which show no signs of agglomeration with time (figure 3c) due to their charged nature unlike the bigger ones (figure 3a). More details of these aging behavior can be found in Section2F. Extended band structures of a solid are crucial [24-26] for inverse Auger processes. This 'hot' $E_3$ exciton can easily avail a large number of final density of states required [10,11] for efficient impact ionization via confinement induced 'zero phonon coupling' to other parts of the PbS band structure in bigger nanocrystallites. This leads to progressive enhancement of spectral broadening of larger (≥16nm) crystallites with decreasing size. Currently, probing any CM of this 'hot' $E_3$ exciton with time domain spectroscopy is beyond the scope due to scarcity of suitable deep-UV pulsed lasers at wavelengths ≤ 210nm. Typically, collisional broadening results in homogeneously broadened

line shapes for gas molecules. However, presence of significant inhomogeneous broadening due to anisotropic exciton-exciton collisions can be explained by size confinement induced momentum uncertainty which connects $E_3$ excitons to different symmetry points in the PbS band structure and a distribution in excitonic K space for such hot excitons.

**E) Mean free path of excitonic collisions and effective size window for efficient multi-excitonic effects.** Unlike monotonic [27-28] size variation of the rates of inverse Auger type processes in smaller nanoparticles, spectral broadening of $E_3$ exciton (figure 1b and 2a) goes through a maximum around intermediate confinement range. This observation can only be explained if impact ionization and subsequent collisional broadening of $E_3$ exciton have a minimum size cut off. Moreover, it was predicted that exciton-exciton scattering [29] and impact ionization [30,31] are one of the main causes of carrier multiplication in semiconductor nanoparticles. However, impact ionization [32] can be suppressed in very small structures due to the presence of quasi ballistic transport. This usually happens when the size (diameter) becomes smaller than the inverse Auger mean free path ($\lambda$) [33] of exciton-exciton scattering. Estimate based on such models [33] yields $\lambda \sim \left(\dfrac{\pi^2 M E_{Ex}}{24 \mu k_B T}\right)^{0.1} (\sqrt{3}) a^*_{Ex} \sim 1.93 a^*_{Ex} \cong 15 nm$ where D=2R~16nm, M≈0.21$m_e^0$, $\mu \approx 0.05 m_e^0$, $a^*_{Ex}$=7.6nm around $E_3$, $k_B$ is Boltzmann constant, T is temperature in Kelvin and $E_{Ex}$~1/R is the binding energy of $E_3$ exciton. Interestingly, this $\lambda$ for 16nm PbS is comparable - (i) to the nanocrystallite diameter at the transition point of both excitonic broadening (figure 2a) and (ii) blue shift (figure 2b) and (iii) the effective Bohr exciton diameter (~15.2 nm) at $E_3$ resonance. Values of this $\lambda$ are plotted in figure 2c. Use of $\varepsilon_\infty$ = 17 produces $\lambda$ >30nm and one hardly expects any PbS nanoparticles having diameters <30nm to undergo MEG/CM in direct contradiction to all experimental reports. However, if we invoke dynamic screening [34,35] of coulomb interaction and $|\varepsilon(\omega)| = 7.2$ is used, then possibility of collisional broadening of $E_3$ exciton within a reasonable *'size window'* of operation is restored inside weak-to-intermediate confinement regimes. This clearly resonates with the theoretical statement[16] that strong confinement may not be exclusively needed for efficient CM. Hence we seriously need to re-examine [34, 35] standard excitonic terminologies to understand the physics of carrier multiplication of such 'hot' excitons – a) having high center-of-mass velocity and b) created inside a region of dispersive dielectric response which can deviate considerably from its usual high frequency limit of $\varepsilon_\infty$. It had also been clearly emphasized [35] that commonplace excitonic parameters like Bohr radius, exciton binding energy can become *"imprecise and ambiguous"* within the framework of Bethe-Salpeter equation which automatically accounts for such dynamical screening. We must however note that, this expression of $\lambda$ is calculated for symmetric S-states only and may need revision in some cases.

**F) Role of dielectric confinement in the aging of absorbance spectra and the observed suppression of aging in smaller nanoparticles.** To get further insights into $E_3$ resonance, aqueous dispersion of PbS nanoparticles of different sizes are deliberately allowed to age and observed variation in optical absorbance are plotted in figure 3. We see that, room temperature $E_3$ excitonic spectra sharpen up with aging (figure 3a and 3b) in bigger nanoparticles. To understand, we briefly focus on the structural development of these nanoparticles during aging. Hydrodynamic size of uncapped PbS was already very large (>2 micron) compared to its mean nanocrystallite diameter of 24nm. The hydrodynamic size of fresh

nanocrystallites with diameter of 16nm was ~30nm, which gradually increases to around a micron due to aging. However, XRD indicates (Supplementary Material; Section II.e) that nanocrystallite size hardly increase during aging, which excludes – (a) significant Ostwald ripening, (b) growth of PbS nanocrystallite by incorporation [36] of sulfur atoms from thiol based ligands. We understand, this increase of hydrodynamic size can come from agglomeration of PbS nanocrystallites which incidentally affect dielectric confinement [37] of these excitons. Overall decrease in absorbance (in spectral range <5eV) of nanoparticles can be explained by steady delocalization of excitonic wavefunction into surrounding material by increased agglomeration and subsequent decrease of oscillator strength. Nonetheless, $E_3$ excitonic resonance is progressively enhanced and sharpened with aging! Spatial delocalization of excitons can reduce confinement induced momentum uncertainty which further reduce the availability of final states within the extended band structure of PbS. Hence, probability of exciton-exciton scattering and collisional broadening of $E_3$ exciton also comes down with time. Here we must point out that even for the biggest PbS nanocrystallite with mean radius 12nm (figure 1b), the spatial extent is well within the weak [13] confinement region $\left(\left(radius \sim 3a_B\right); a_{Ex}^{*} = 7.6nm\right)$ for all these quantum effects to influence the excitonic delocalization. Additional defect formation or incorporation of water [38] into the nanocrystallites cannot explain such systematic but seemingly opposite aging dependence of – (a) the $E_3$ excitonic resonance and (b) the absorption spectra in lower energy ranges. Similar effect, although less pronounced, is observed in 16nm PbS (figure 3b). Here $E_3$ exciton showed initial red shifts (figure 3b), during aging induced gradual lessening of dielectric confinement, which was absent in larger uncapped samples due to its quasi-bulk nature. Unlike 16nm PbS, hydrodynamic diameter and absorbance of smaller 3nm PbS are hardly changing with time (figure 3c) This can only happen if these smaller nanoparticles are intrinsically charged as compared to the bigger ones. Mutual coulomb repulsion of charged nanoparticles can prevent agglomeration, sustain the dielectric environment of 3nm PbS and safeguard the absorbance from aging effect.

**G) Suppression of auger cooling in smaller nanoparticles.** As the $E_3$ resonance ~5.9eV is fading out with increasing temperature from 20°C to 100°C for uncapped quasi-bulk PbS (figure 4a), another peak slowly emerges ~5.3 eV at its expense. Interestingly, 5.3eV is also the bulk $E_3$ critical point of PbS! We see no gradual energy shift of the $E_3$ resonance with increasing temperature but rather a coexistence of $E_3$ resonance and the bulk like feature from 80°C to 95°C. Above that temperature, $E_3$ resonance vanishes altogether. However, we again retrieve this $E_3$ resonance once temperature is reduced back to 20°C, except for some aging induced reduction in absorbance and sharpening of the $E_3$ peak as discussed in the previous section. PbS has a positive temperature coefficient of band gap except under strong confinement regime where excitonic energy hardly [2] varied with temperature. However, we cannot use the same reasoning to explain current observations for this weakly confined quasi-bulk PbS! With increasing phonon scattering and decreasing $\Delta T = (T_{Hot\ Carrier} - T_{Lattice})$ above room temperature, this $E_3$ exciton has much less access to other parts of PbS band structure for efficient impact ionization. Therefore, emergence of bulk like spectral feature ~5.3eV with increasing temperature represent Auger cooling [39,40] of energetic 'hot' $E_3$ excitons to its bulk $E_3$ band edge by transferring its excess energy mostly to heavier holes ($m_e^* \neq m_h^*$ at $E_3$ [14]). Prominence of bulk $E_3$ band edge steadily decrease with decreasing size because of stronger size confinement and consequently lesser density of available ground states at smaller sizes (figure 4b). Observed agglomeration (figure 3a and 3b) of these bigger nanocrystallites (D≥16nm) also indicates their non-charged nature which may have prevented the reported [41] the suppression of Auger cooling seen in smaller charged nanoparticles (figure 3c & 4c). Nanocrystallite diameter also remains same (Supplementary Material; Section II.h) which confirms the absence of significant irreversible changes like oxide/defect formation in this range of temperatures. Especially, line width broadening of $E_3$ exciton with decreasing size and with increasing temperature are qualitatively different. In fact, the temperature (T) coefficient ($TC_{II}$) of impact ionization depends on its threshold energy ($E_i$) and the mean free path ($\lambda$) of exciton scattering as [42] $TC_{II} = \frac{E_i}{q\varepsilon\lambda}\left[\frac{d\lambda}{\lambda dT} - \frac{dE_i}{E_i dT}\right]$. Spectra of 3nm PbS nanoparticles follows the literature [43] for annealing temperatures >40°C, where it undergoes Ostwald ripening and we also see that hydrodynamic size of these particles increases to >275nm. However, the $E_3$ resonance did not change till 30°C. This apparent temperature independence of excitonic line width for strongly confined PbS can be ascribed to a set of reasons like – (a) reduction of impact ionization due to charged nature of these strongly confined particles, (b) vanishing of $TC_{II}$ due to mutual cancellation of the above two terms, (c) reduction of collisional broadening and Auger cooling in the ballistic regime, (d) strong surface effects etc. Size dependence of photocharging [44] and/or photoinduced surface trapping [45,46] are currently being investigated.

Recently, it was argued [47] that higher energy excitonic states are bulk like where Bohr exciton radius ($a_B$) can approach to 'zero' value due to the divergence of effective mass. However, such 'arithmetical' divergence of effective mass happens $\left(d^2 E_{C,V}/dk^2 = 0\right)$ away from $E_3$ like critical points $\left[\nabla_k (E_C - E_V) = 0\right]$. It is also evident [48] that charged particles with smaller effective masses are easy to accelerate and classical mechanical treatment of effective mass approximation no longer holds where $m^*_{e,h} = \pm\hbar^2/\left(d^2 E_{C,V}/dk^2\right)$ starts to diverge.

## 3. CONCLUSIONS.

In summary, dynamical coulomb screening and the residual effect of extended band structure of PbS were sought to explain the dependence of $E_3$ excitonic resonance with aging and temperature over a wide range nanoparticle sizes from weakly confined quasi bulk to strongly confined very small nanocrystallites. We argue that Bohr radius of hot exciton is not material specific but photo-excitation specific [49]. This empirical notion of effective excitonic Bohr radius is a direct consequence of the dynamical screening of coulomb interactions at high photon energies which may eventually lead to better conceptual understanding of condensed matter physics of 'hot' excitons. Although it is not immediately clear whether one can use such a straight forward extension of hydrogenic atom model to understand the formation of excitons at such high enough energies. However, it is understandable that significant reduction of carrier multiplication in the ballistic limit could have influenced all past studies focused only on strongly confined quantum dots. Therefore, we also predict an intermediate *'size window'*, where impact ionization can dominate over other kinds of exciton relaxation pathways. Although, this current 6eV excitonic resonance has no relevance for photovoltaics, but one can surely extend these analyses of hot exciton physics and even suggest size optimization of semiconductor nanoparticles for better exploitation of carrier multiplications to improve the power conversion efficiency of nanophotovoltaic cells inside the solar spectrum. Recently, it has been brought to our notice that enhancement of internal quantum efficiency of PbSe quantum dots with increasing size is verified [50] in experiements.

**Acknowledgements:** Authors want to thank Dept. of Science and Technology, India for DST Nano Unit grant SR/NM/NS-42/2009 and IISER-Pune for support. SD wishes to thank Prof K. N. Ganesh and IISER-Pune for startup funding.

# FIGURE 1

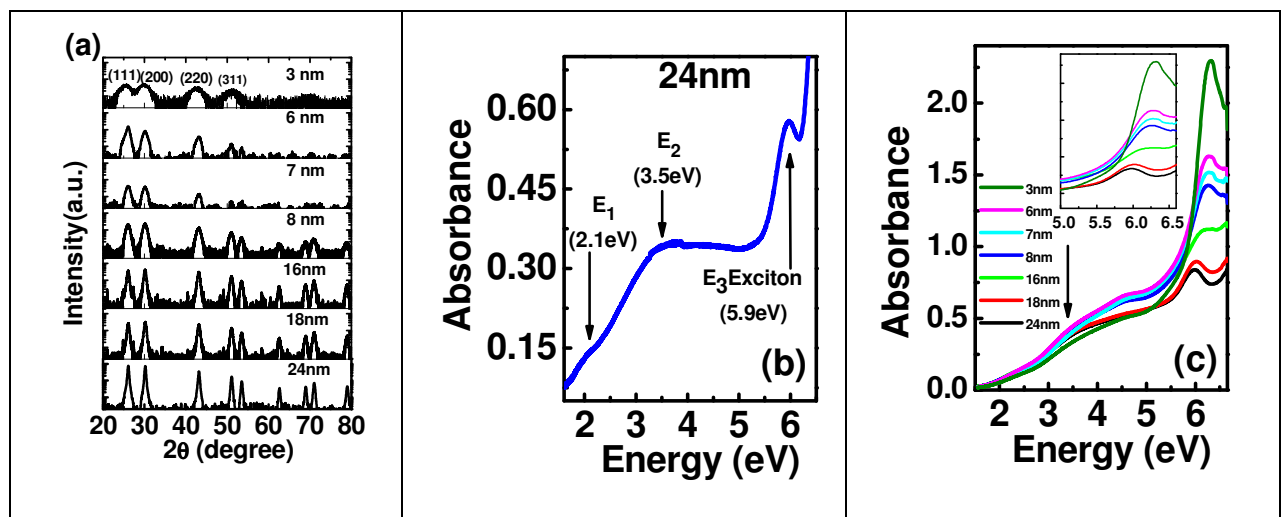

**Figure 1:** **(a)** XRD peaks broadens with decreasing crystallite sizes (diameters). **(b)** Room temperature optical absorption spectra of freshly prepared un-capped PbS nanocrystallites. The term 'freshly prepared' means the time (t = 0) starts ~60 mins after the end of each reaction. The $E_3$ transition of these large PbS nanoparticles without TG capping cannot be from any unused Thio-glycerol present in solution. **(c)** This plot demonstrates the enhancement of excitonic absorbance with decreasing size of PbS nanocrystallites. Molar concentrations of PbS were kept same for all crystallite sizes during the measurements except for 3nm particles. There it was kept at 50% of the rest to avoid the saturation of photo-detector. The inset shows close up of the $E_3$ excitonic peak (> 5.0 eV). We also notice that the $E_3$ excitonic spectral shapes for sizes smaller than 16nm are not at all symmetric unlike bigger particles. Therefore any '*single*' Gaussian or Lorentzian or even a Voigt type line shape is not a good fit to these peaks. Size variation of $E_3$ excitonic resonance can only arise from solid PbS and not from any residual chemical.

# FIGURE 2

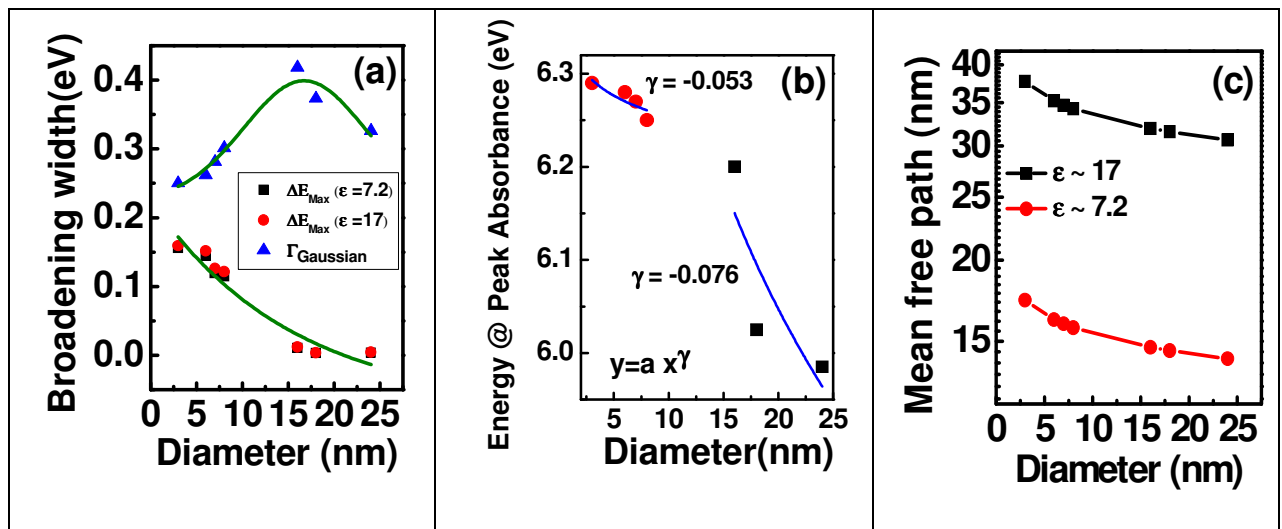

**Figure 2: (a)** De-convoluted Gaussian component of the $E_3$ excitonic line width is largest around 16nm. Maximum possible energy broadening ($\Delta E_{max}$) due to the size distribution is much smaller than both the homogeneous and inhomogeneous line widths for all sizes. **(b)** This plot shows monotonic blue shifts of $E_3$ excitonic resonance (5.3eV + $\Delta E$) with decreasing mean crystallite diameter of PbS nanoparticles. Notably, no single power law behavior can describe the variation for all sizes. **(c)** A comparison of exciton mean free path values calculated using $\varepsilon_\infty = 17$ and $|\varepsilon(\omega)| = 7.2$. Calculated mean free path for exciton-exciton scattering event is >30nm for the usual $\varepsilon_\infty = 17$. Solid lines are just a guide to the eye only.

# FIGURE 3

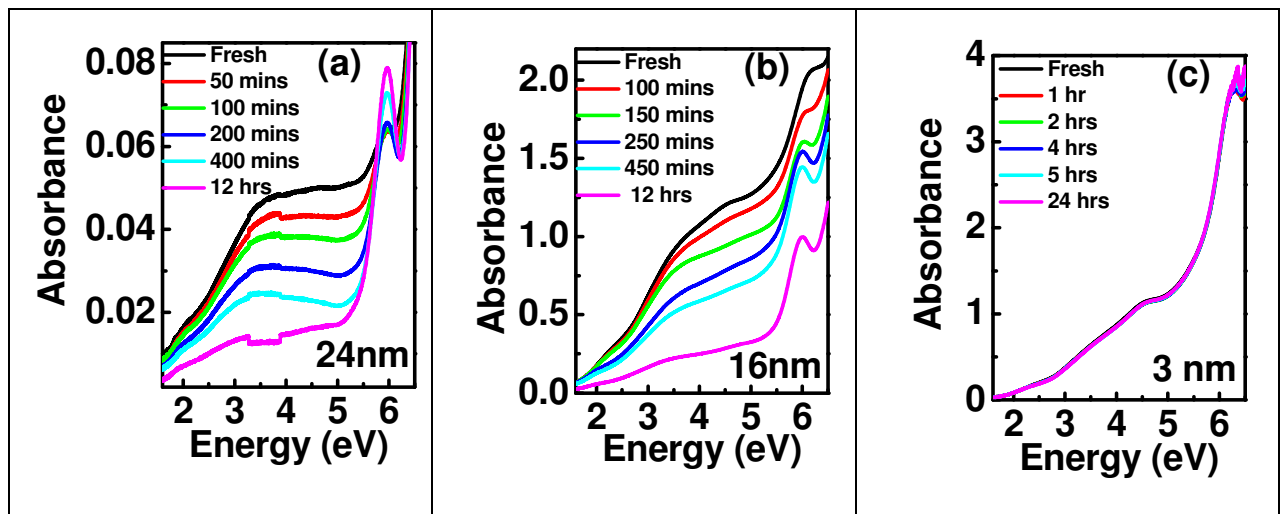

**Figure 3:** Aqueous dispersion of PbS nanoparticles were not sonicated and deliberately allowed to age to get better physical insights of the spectral origin of $E_3$ exciton. **(a)** Variation of absorbance of $E_3$ exciton is characteristically opposite to the portion of the spectra at photon energies lower than 5eV for uncapped PbS nanocrystallites with mean diameter of 24nm. The abrupt changes between 3eV to 4eV are due to instrumental artifacts for lamp changes etc at small absorbance. **(b)** We see qualitatively similar kind of sharpening of $E_3$ excitonic peak with aging for 16nm PbS. **(c)** The absorption spectra for strongly confined PbS nanoparticle with mean diameter 3 nm hardly changes with aging as compared to that of figure 3a and 3b. Most likely cause of this is the charging behavior of smaller nanoparticles.

# FIGURE 4

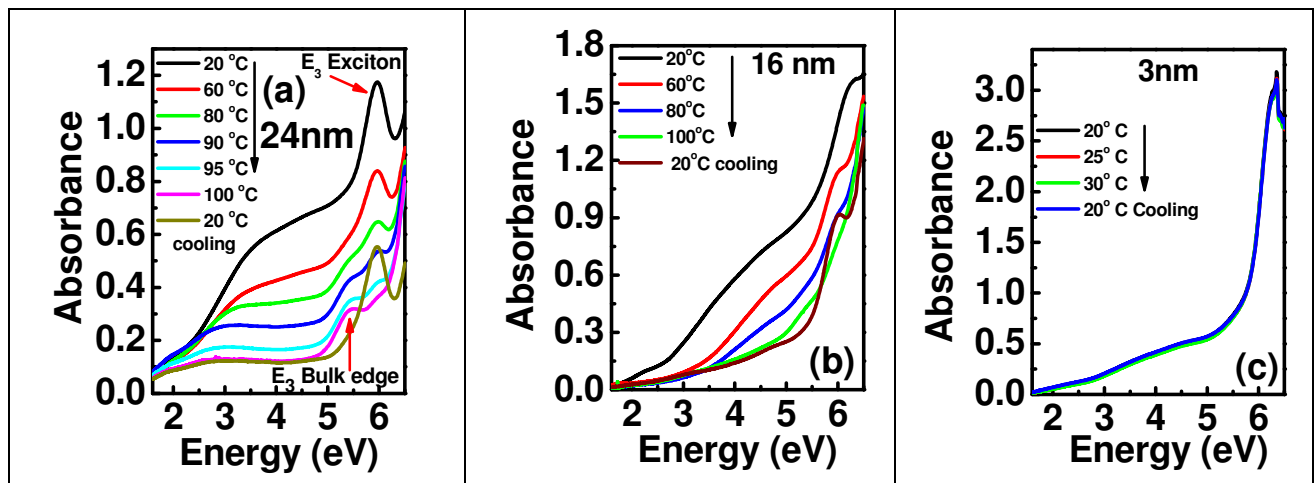

**Figure 4: (a)** Temperature variation of uncapped PbS nanocrystallite with mean diameter of 24nm. The $E_3$ bulk edge (~5.3 eV) appears at the expense of $E_3$ exciton and 'coexist' for temperatures >80°C. $E_3$ excitonic feature nearly vanishes by 100°C but reversibly recovers around 20°C during cooling. **(b)** Almost similar but less pronounced spectral changes with temperature are also observed for 16nm PbS during cooling stages. **(c)** 3 nm PbS shows relative less temperature dependence upto 30°C. Arrows indicate the direction of temperature variation in all three graphs. These results fully compliment the behavior shown in figure 3 as the charged nature of smaller nanoparticles not only prevent agglomeration but also suppress Auger cooling.

: SUPPLEMETAL MATERIAL :

# Title: Optical Studies of Hot Excitons in Lead Sulfide Nanocrytallites

**Authors:** *Padmashri V. Patil\* and Shouvik Datta\*\**

**Affiliations:** Department of Physics, Indian Institute of Science Education and Research-Pune,

1st Floor, Central Tower, Sai Trinity Building, Pashan, Pune – 411021, Maharashtra, India.

Email : *padmashriv.patil@gmail.com, **shouvik@iiserpune.ac.in,

**I) EXPERIMENTAL METHODS: Synthesis, Purification and Characterizations Tools.**

Chemicals used in the wet chemical synthesis of PbS nanocrystallites were Lead Acetate $Pb(CH_3COO)_2$ and Sodium Sulfide ($Na_2S$). Thio-glycerol (3-Mercapto-propane-1,2diol or TG) was used as capping agent. All chemicals were purchased from Sigma Aldrich and used without further purifications. Synthesis was done in a two neck, round bottom flask with 18MΩ de-ionized water as solvent. 10mL of 0.02 M Lead Acetate was initially heated to $80^o$ C. Then another 10mL of 0.02 M Sodium Sulfide and TG was added drop wise to hot lead acetate. The mixture was stirred continuously at $80^o$ C for half an hour. Quantity of TG was varied during the reaction (from none for the uncapped sample to 120μL in steps of 20μL) to get a wide range of nanocrystallite sizes. After the reaction, particles were first washed with de-ionized water and then centrifuged for 10 minutes. This sequence of washing and purification was repeated for a total of 5 times. Finally, these are re-dispersed in de-ionized water for further optical characterizations. 20μL of this final aqueous dispersion of PbS nanoparticles is then mixed with 3mL of de-ionized water inside a quartz cuvette for optical absorption studies.

Perkin Elmer's Lambda 950 was used for UV-VIS-NIR optical absorption spectroscopy with scan steps of 0.2nm. Crystallite sizes were determined by X-ray powder diffraction (XRD) spectra of drop casted thin films of nanoparticles using Bruker D8 Advanced X-ray diffractometer with Cu-K-α X-ray radiation at a wavelength of 0.154 nm and increment of $0.01^°$ per step. Moreover, hydrodynamic sizes were also determined by Dynamic Light Scattering (DLS) studies using Malvern Zeta-Sizer Nano-ZS90. DLS experiments were carried out at 25°C using 633nm, 3mW He-Ne laser using 90° optics having 10μm apertures.

## II) Crystalline Quality and Size determination of PbS Nano-Crystals.

### II.a) Determination of PbS Nanocrystallite sizes from XRD spectra.

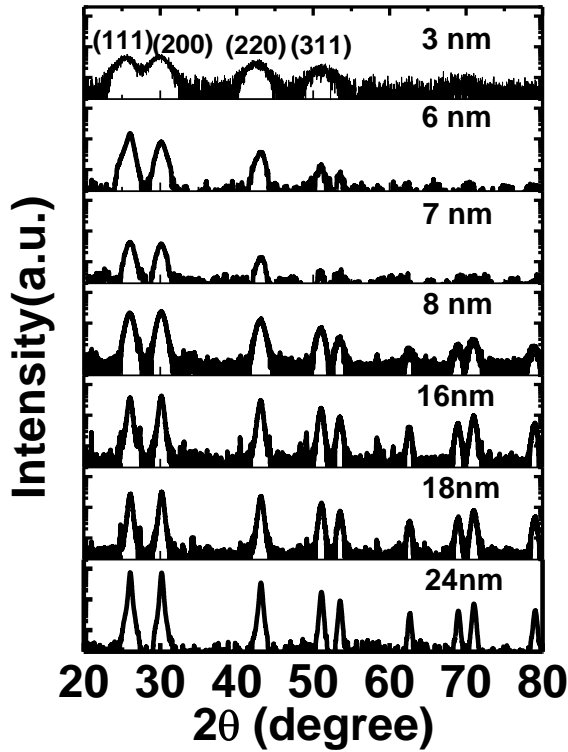

Figure II.a (also figure 1a in the manuscript) shows the XRD spectra for PbS nanoparticles of various sizes. The 2θ peak positions at $26.0°$, $30.1°$, $43.1°$, and $51.1°$ can be indexed respectively to planes (111), (200), (220), and (311) of cubic rocksalt structure of PbS. Each of these peaks are used to calculate the final mean size of PbS nanocrystallites using the Debye-Scherrer formula $D = \dfrac{K\lambda_{XR}}{\beta \cos\theta}$, where D is the average size(diameter) of the crystallite, K is the shape factor $\approx 0.9$, wavelength $\lambda_{XR} = 0.154$ nm for Cu-Kα X-Ray, β is the FWHM of the Bragg peaks and θ is the Bragg angle. Estimated mean crystallite sizes (diameters) are also mentioned in figure. Clearly the width of the Bragg peaks increases with decreasing crystallite sizes. We subsequently use these mean crystallite sizes in our study. Further analysis of the powder diffraction spectra using Williamson-Hall method showed very little strain (<0.05%) in these crystallites, which substantiate the validity of our calculation based on Debye-Scherrer formula.

### II.b) Estimated band gap values for different crystallite sizes of PbS nanoparticles.

| Crystallite Diameter (nm) | Calculated Bulk band gap ($E_0$) in Wavelength (nm) Unit |
|---|---|
| 24 | 2853 |
| 18 | 2587 |
| 16 | 2451 |
| 8 | 1417 |
| 7 | 1211 |
| 6 | 991 |
| 3 | 327 |

Calculated values of band gap ($E_0$) absorption wavelengths for different crystallite sizes. Values are calculated by using effective mass approximation formula

$$\Delta E = E - E_g = \frac{h^2}{8R^2}\left[\frac{1}{m_e^*} + \frac{1}{m_h^*}\right] - \frac{1.8e^2}{4\pi\varepsilon R}$$

where $E_g$ is bulk band gap energy, E is band gap energy due to confinement, $\varepsilon$ is the optical frequency dielectric constant of the material, e is electronic charge, h is the Plank's constant, $m_e^*$ and $m_h^*$ are effective mass of electron and holes respectively.

### II.c) Absorption spectra of $E_0$ transitions in PbS nanoparticles of various sizes.

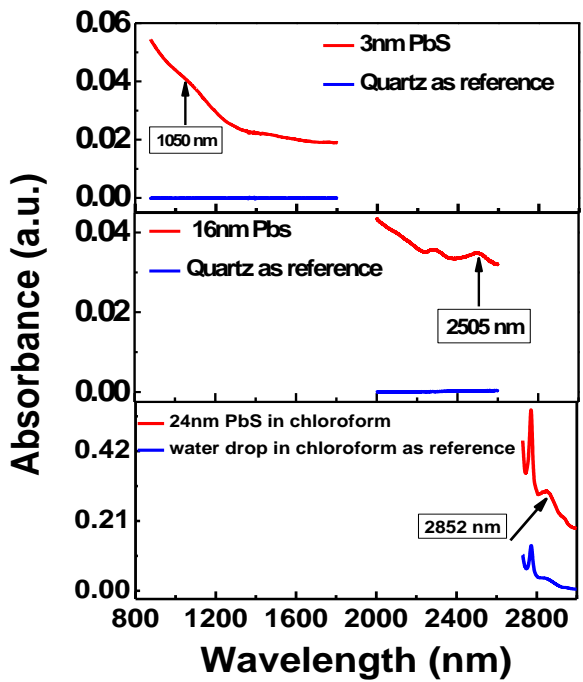

Figure II.c shows $E_0$ optical absorption spectrum of freshly prepared PbS nanocrystallites of mean diameter ~3nm and 16 nm in thin film and in chloroform for 24nm PbS nanocrystallites at room temperature. Observed values of $E_0$ for different sizes are marked in the respective absorbance spectra. For 3nm PbS the observed absorption wavelength of $E_0$ transition is 1050nm; which is far more than what we get by using effective mass approximation. This shift for very small particles may be due to approximation fails for smaller particles.

We have difficulties in recording the spectra of 24nm PbS crystallites in thin film form because of strong water absorbance in this (>2000nm) wavelength region. We even tried to record spectra of 24nm PbS dispersed in chloroform by repeated use of centrifuge and vacuum drying. However, we found that it is not easy to completely remove traces of water from these samples. Therefore, we suspect that small quantity of water molecules trapped inside cages of agglomerated clusters of PbS nanoparticles produce the observed absorption peak around 2850nm. Magnitude of this absorbance also grows with the addition of water drops as well as with the addition of PbS nano-particles. Therefore, the expected $E_0$ transition in 24nm PbS may be buried under this strong water absorption peak.

**II.d) TEM image for 24 nm PbS nanoparticles: Showing larger particles agglomerate to form micron sized clusters.**

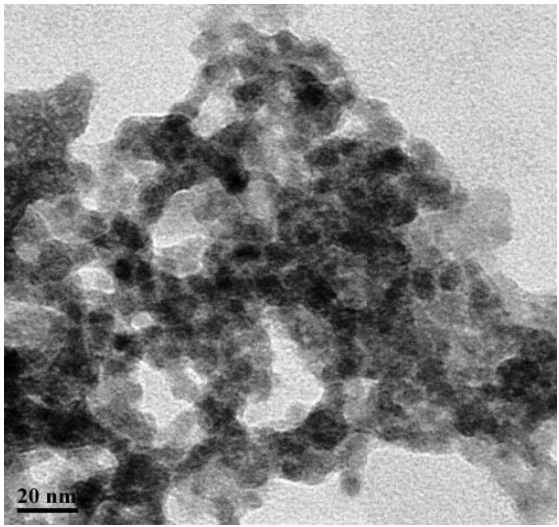

Figure II.d shows TEM image for 24nm PbS nanoparticles. TEM image was taken with aged uncapped PbS nano-particles which were not ultra-sonicated deliberately to reveal its intrinsic agglomeration. The image shows granular cluster formation of for PbS nanocrystallites.

**II.e) XRD spectrum – Aging does not change nanocrystallite sizes.**

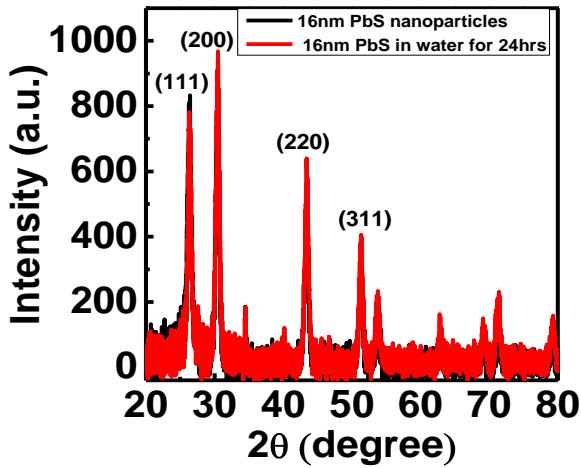

Figure II.e shows XRD spectrum for 16nm PbS nanocrystallites. One spectrum is recorded on freshly prepared nanoparticles and for other spectra sample is intentionally kept in water for 24 hrs. Drop casted thin films of each sample are used for measurements. It shows that aging doesn't changes crystallite size of ~16nm PbS nanoparticles. So, all changes observed in UV-VIS absorption (figure 3b in manuscript) of 16nm PbS aging are due to agglomeration of nanocrystallites and not due to Ostwald kind of growth of these nanocrystallites.

**II.f) DLS results on aging of 16nm PbS nanoparticles - Aging certainly affects agglomeration and increases the hydrodynamic sizes of the nanoparticles.**

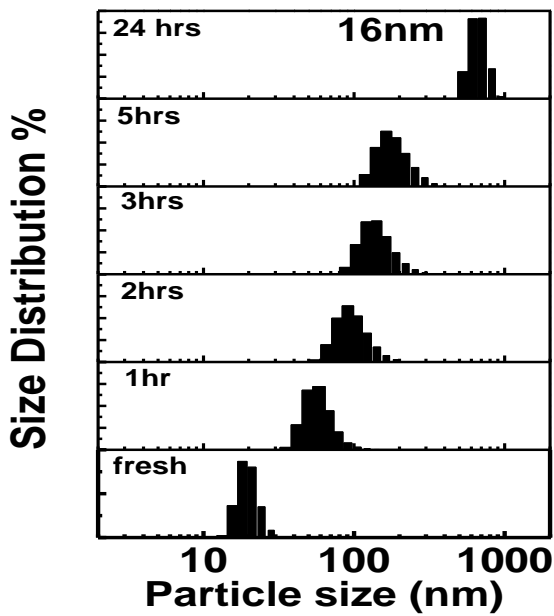

Figure II.f shows effect of aging on hydrodynamic size of 16nm nanocrystallites. It is observed that starting with mean hydrodynamic size ~16nm nanoparticles size increases to micron size due to Van der Waals type of attractive interaction.

This figure II.f along with figure II.e, indicates that increase of hydrodynamic size is due to the agglomeration of nanocrystallites (figure 3b in manuscript) via Van der Waals type of attractive interaction only and not by Ostwald kind of ripening process where bigger nanocrystallite grow at the cost of smaller ones.

**II.g) DLS data on aging of 3nm PbS nanoparticles – Hydrodynamic size do not change in smaller nanparticles within the experimental accuracy.**

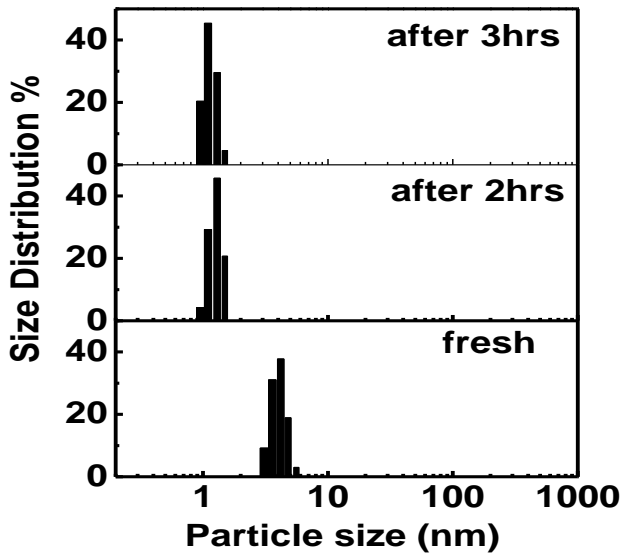

Figure II.g shows the effect of aging on hydrodynamic size of ~3nm PbS nanoparticles. Strongly confined nanoparticles are expected to be charged and show no significant aggregation (fig 3c in manuscript). Understandably, DLS estimates of these very small particles are inaccurate by around ± few nm. **But it is very clear that unlike 16nm PbS shown in figure II.f, the hydrodynamic size of 3nm nanocrystallites is not increasing to micron level.**

**II.h) XRD spectrum showing no significant effect of temperature on crystallite size.**

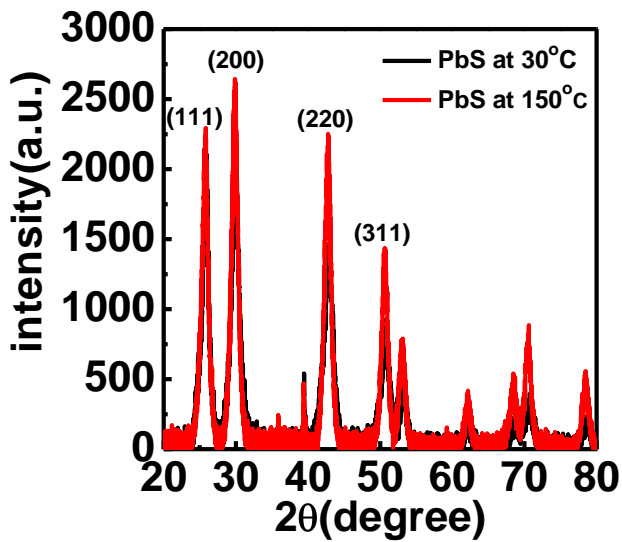

Figure II.h shows XRD spectrum of ~8nm PbS nanocrystallites at room temperature and at 150°C. It shows heating of PbS nanoparticles upto 150°C doesn't significantly changes the spectral broadening and the crystallite size of these nanoparticles remains unchanged. So the temperature dependence spectral change reported in figure 4 of the manuscript is not due to change in crystallite size by oxide formation etc.

## II. i) Effect of confinement on kinetic energy as well as on binding energy.

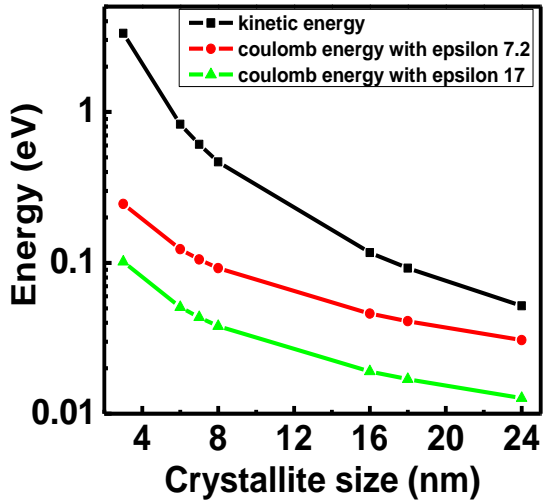

Figure II.i shows effect of confinement on kinetic energy $\dfrac{h^2}{8R^2}\left[\dfrac{1}{m_e^*}+\dfrac{1}{m_h^*}\right]$ of electron and hole as well as on coulomb interaction energy $\left|-\dfrac{1.8e^2}{4\pi\varepsilon R}\right|$ of electron and hole. Energies are calculated by using effective mass approximation formula. The observed $E_3$ excitonic features at 5.9eV has shift from its $E_3$ by 0.6eV, this much high energy shift cannot be explained by effective mass approximation theory alone except for < 3nm particles.

## III) Calculated exciton binding energy for PbS nanoparticles using ε = 17 and ε = 7.2

**Enhancement of binding energy at smaller |ε| for $E_3$ resonance.**

| Samples ; Nano-Crystallite Diameter (nm) | Exciton Binding Energy $E_{Ex}$ (eV) assuming $\varepsilon_\infty = 17$ | Exciton Binding Energy $E_{Ex}$ (eV) assuming $|\varepsilon| = 7.2$ |
|---|---|---|
| NO TG PbS ; 24 nm | 0.013 | 0.030 |
| 20 TG PbS ; 18 nm | 0.017 | 0.040 |
| 40 TG PbS ; 16 nm | 0.020 | 0.045 |
| 60 TG PbS ; 8 nm | 0.038 | 0.090 |
| 80 TG PbS ; 7 nm | 0.044 | 0.103 |
| 100 TG PbS ; 6 nm | 0.051 | 0.120 |
| 120 TG PbS ; 3 nm | 0.102 | 0.240 |

# IV) Control experiments to rule out presence of artifacts and unused chemicals or ligands in the optical measurements of $E_3$ transition.

## IV.a) Water absorption spectrum- No sign of $E_3$ resonance peak around 5.9-6eV region.

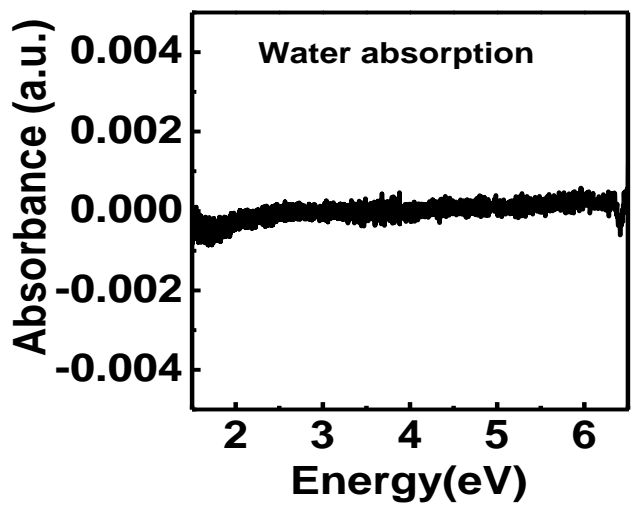

Figure IV.a shows absorption Spectra of de-ionized water with water as reference. No trace of PbS like $E_3$ peak found around 5.9 eV. This also indicates that reported $E_3$ transition of PbS are not due to any instrumental artifact or defects in cuvette or from water itself.

## IV.b) Absorption spectrum of CdTe nanoparticles – Clear absence of $E_3$ like transitions around 6eV as seen in PbS nanoparticles.

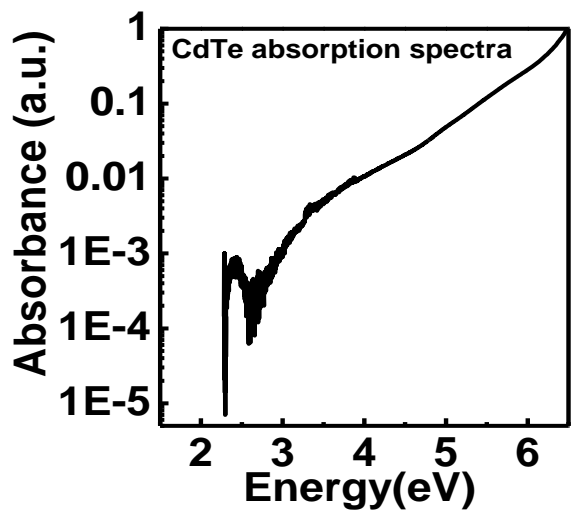

Figure IV.b shows, absorption Spectra of CdTe nano-particles (capped with Mercapto Succinic Acid) with de-ionized water as reference. No trace of PbS like $E_3$ resonance peak found around 5.9 eV.

This also indicates that reported $E_3$ transition of PbS are not due to any instrumental artifact or defects in cuvette.

**IV.c) Absorption spectra of thioglycerol capping agent - Absence of any resonance peak around or below 6eV.**

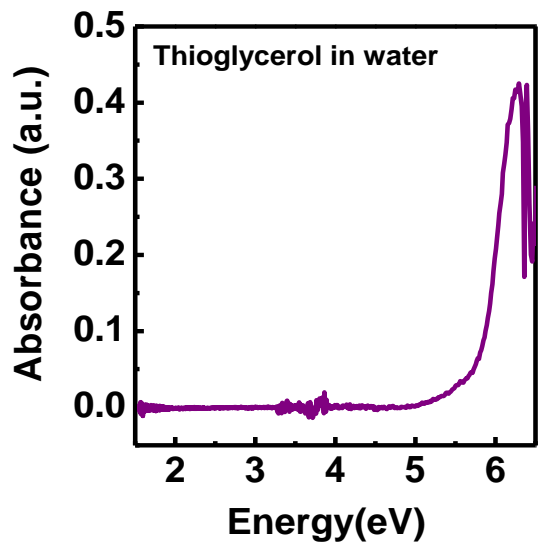

Figure IV.c shows absorption spectra of Thio-glycerol solution in de-ionized water with deionized water as reference. It shows no resonance transition around 5.9 eV. **Also the absorption spectra for 24 nm uncapped nanoparticles presented in the paper (Figure1a, 3a, 4a) have no thioglycerol in it.** Therefore, the $E_3$ transition for PbS nanoparticles (reported in manuscript) cannot be from any unused Thio-glycerol (i.e. ligand) present in solution.

**IV.d) Absorption spectra of Sodium Sulfide.**

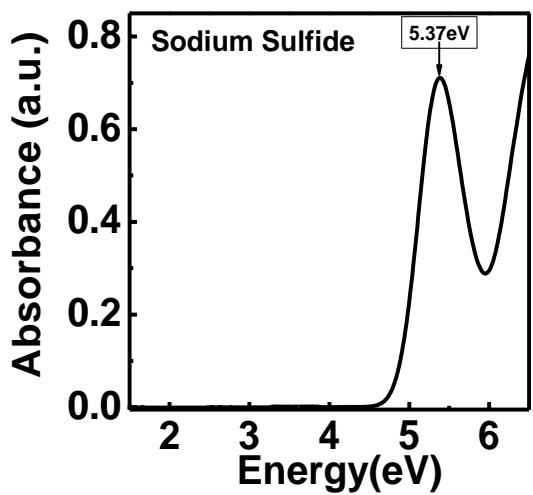

Figure IV.d shows absorbance of aqueous Sodium sulfide with deionized water as reference. It shows absorption at 5.37eV, indicating that $E_3$ resonant absorbance of PbS nanoparticles (>5.9 eV) is not due to presence of unused sodium sulfide in the aquous dispersion of PbS nanoparticle used in optical absorption experiements.

**IV.e) E$_3$ resonance is not coming from unused or left over Lead Acetate – Evidence that our washing and purification sequence removes all leftover reagents.**

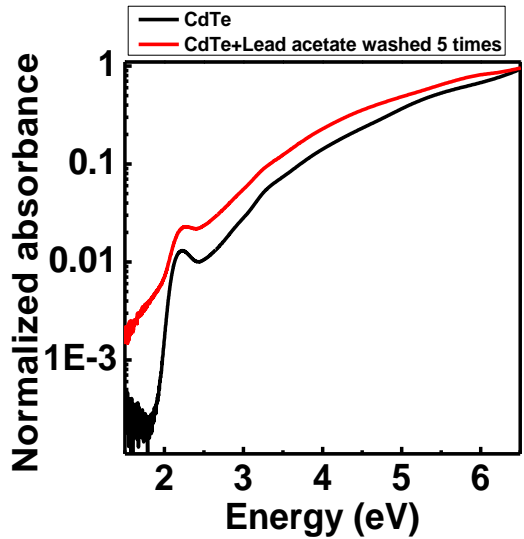

**Evidence that E$_3$ excitonic resonance is not coming from any leftover chemical reagents as our washing and purification sequence removes it from the final aqueous dispersion of PbS nanoparticles used in optical absorption experiments.**

Figure IV.e shows absorbance spectra for mercaptosuccinic acid capped CdTe nanoparticles and CdTe + additional Lead acetate solution, washed for 5 times. Here we have done the same experiment with CdTe nanoparticles which has no E$_3$ like resonant peak to beginning with. We added aqueous lead acetate solution (20 microliter of 0.02M lead acetate in 3ml of CdTe solution, this is the full amount we add during the chemical synthesis in aqueous dispersion of CdTe and then we washed it 5 time using centrifuge to prove that unused lead acetate won't persist the washing + purification sequence. Moreover, in case of PbS nanoparticles, >99% of this 20 micro-liter of lead acetate is expected to be used up in making PbS and only a very small fraction will be left after the reaction with equal concentration of Na$_2$S. **So the E$_3$ resonance peak is definitely not coming from unused Lead Acetate either.**

**IV.f) Lead sulfide nanoparticles synthesized using lead acetate and H$_2$S is showing the prominent E$_3$ resonance peak.**

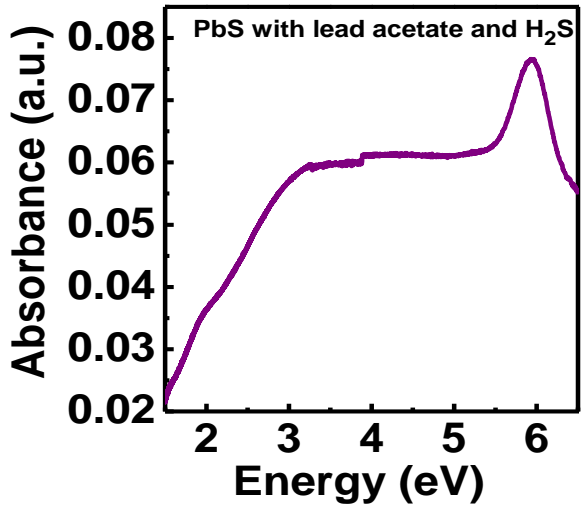

Figure IV.f shows similar spectral features of excitonic transitions (figure 1a in manuscript) are also reproduced in PbS nanoparticles prepared with lead acetate and H$_2$S gas. So this experiment also proves that the E3 transition is not coming from any artifact or presence of unused chemicals (sodium sulfide).

**IV.g) Absorption spectra for 45nm PbSe nanocrystallites – Not showing any resonance peak at 5.9-6 eV.**

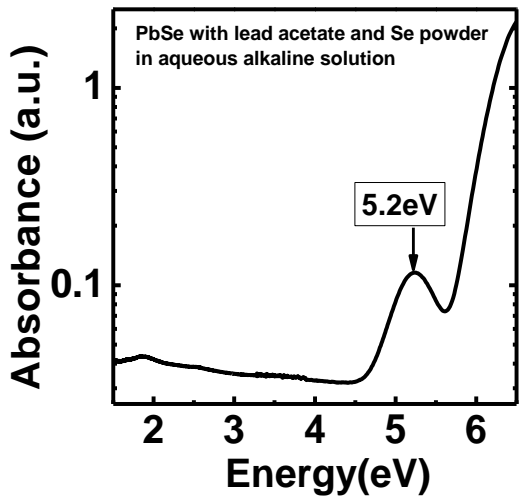

Figure IV.g shows absorption Spectra of uncapped PbSe nano-particles with de-ionized water as reference. No trace of PbS like E$_3$ peak is found around 5.9 eV. For bulk PbSe E$_3$ occurs 4.5eV, here it is observed as blue shifted might be due to confinement effect.

Importantly, the E$_3$ transition of PbSe is around 5.2eV which is at a much smaller energy than 5.9 eV; this also proves that reported E$_3$ transition of PbS are not due to any instrumental artifact or defects in cuvette but it is material's property.

# V) Control experiments to rule out any artifacts from temperature dependent and time dependent measurements.

## V.a) Aging of Lead Acetate do not account for aging induced enhancement of $E_3$ resonance as observed in PbS (figure 3a).

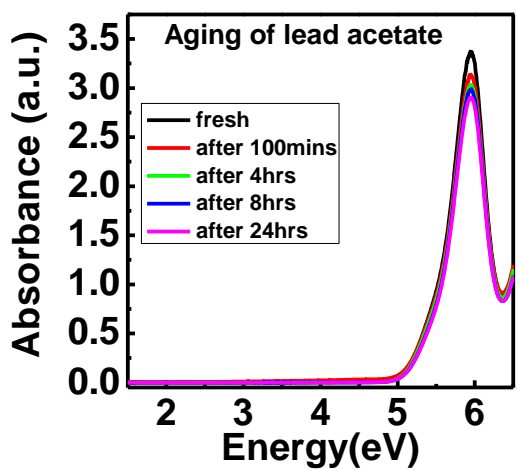

Figure V.a shows aging of aqueous lead acetate with different time interval. Spectra taken by dissolving lead acetate in water with water as reference. Peak feature at 5.9eV degrades in intensity with time which is just opposite to behavior for uncapped PbS nanoparticles. It shows the feature observed in aging of uncapped PbS nanoparticles (Fig 3a in manuscript) is not coming from presence of unused lead acetate in the final product of PbS nanoparticles.

## V.b) Lead Acetate absorption spectra at different temperature.

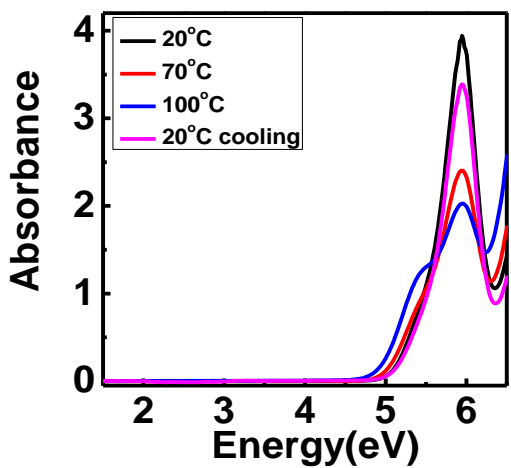

However, we will clearly demonstrate in the next section (Figure VI.a and Fig 1b & c in the manuscript) that the $E_3$ resonance is not coming from the leftover Lead Acetate precursors. Moreover, we have already shown in figure IV.e that our washing and purification sequence remove all such leftover reagents.

Figure V.b shows effect of temperature on absorption spectra of aqueous lead acetate solution. Spectra taken by dissolving lead acetate in water with water as reference. With increasing temperature absorption peak at 5.9eV fades out and peak at 5.4eV start appearing, but lowering temperature back to 20°C shows peak at 5.9eV with no signature at 5.4eV. This behavior is similar to the uncapped PbS nanoparticles. Similar spectral behavior and temperature dependence (with figure 4a in manuscript) just indicates that the molecular orbitals of $E_3$ band structure may be generically related with atomic orbitals of lead in lead compounds.

## VI) Temperature dependence and Aging of Optical Absorption Spectra of uncapped PbSe nanoparticles and its comparison with uncapped PbS nanoparticles.

At this stage, we want to mention some points which makes PbS different than PbSe are dielectric constant ($\varepsilon_\infty$) of PbS is 17 and that for PbSe is 25. Moreover, PbS ($E_g \sim 0.4$eV) also has a larger band gap than PbSe ($E_g \sim 0.3$eV). As a result, the Bohr radius for PbS ($a_B=18$nm) is smaller than that for PbSe ($a_B=45$nm). Therefore, PbSe having quite high dielectric constant and smaller band gap can have very low excitonic binding energy as well as very high coulomb screening effect as compared to PbS. We have observed the manifestation of these differences in the following supplementary plot VI.a.

### VI.a Temperature dependent study for uncapped 45nm PbSe Nanocrystallites.

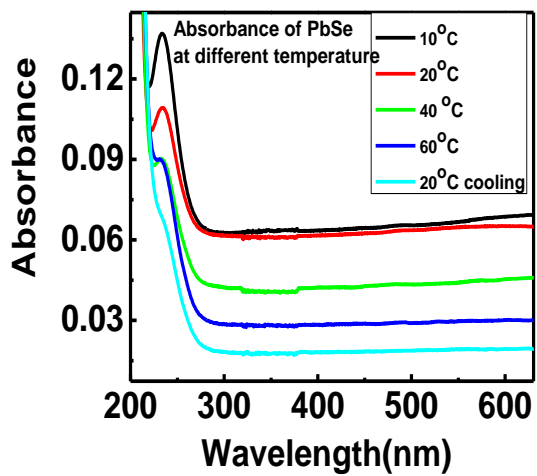

PbSe is synthesized using Lead acetate and Se powder in alkaline solution. After synthesis we washed it several times with de-ionized water and finally dispersed in de-ionized water to check UV-VIS absorption.

The $E_3$ absorption peak of PbSe is ~ 233nm (5.3eV) not ~210nm (5.9eV as shown in figure 1b in manuscript). Moreover, the observed $E_3$ transition (233nm) of PbSe around room temperature is already around the same place of the calculated $M_1$ type $E_3$ transition (225nm, $\Sigma_4 \rightarrow \Sigma_7$) [Ref. 14; Phys. Rev. B **8,** 1477 (1973)]. So it seems that under the current operating temperatures the $E_3$ transition of PbSe is already in the Auger Cooled state due to its smaller binding energy as compared to PbS. The same explanation applies for the observed irreversible change in the absorbance too.

**Sizeable difference in the transition energy of $E_3$ resonance for quasi-bulk PbS (210nm) and that of PbSe (233nm) also strongly suggests that any leftover lead acetate is not contributing the $E_3$ resonance.**

**VI.b Aging of absorption spectra for 45nm PbSe Nanoparticles.**

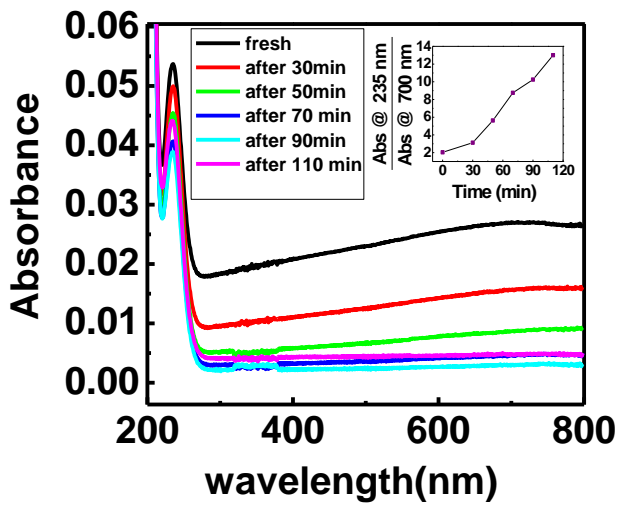

**VI.c Aging of absorption spectra (fig 3a) for 24nm PbS Nanoparticles is provided for comparison.**

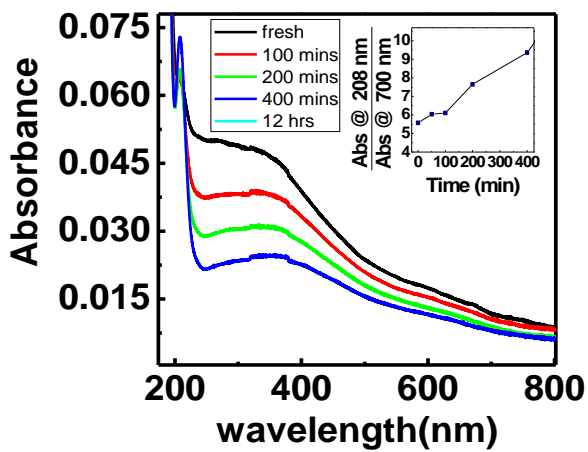

**Evidence of enhancement of $E_3$ absorbance in both PbS and PbSe :**

Figure VI.b shows aging study for PbSe nano-particles, dispersed in water. We compared aging behavior bulk PbSe (45nm) and bulk PbS (24nm) (Figure VI.c and figure 3a in the manuscript). We deliberately allowed these nano-particles to age and absorption spectra were taken at different stages of aging. Insets show ratio of absorbance value at 235nm to absorbance value at 700nm for 45nm PbSe and ratio of absorbance at 208nm to absorbance value at 700nm for 24nm PbS.

This ratio increases with aging duration in both nano-particles. Therefore, we demonstrate that the absorbance of $E_3$ transition in both PbS and PbSe enhances with time as compared to the absorbance at lower photo-excitation energy. This is in line with our explanations given in subsection 2F in the manuscript of the aging process of $E_3$ resonance.

**VII. Future plan on time resolved studies in deep-UV range (< 250nm) to probe the femto-second dynamics of hot $E_3$ exciton.**

We plan to study transient absorption measurements for $E_3$ transitions in PbS and PbSe nanoparticles (in collaboration with the THz-Spectroscop Laboratory in IISER-Pune. TOPAS Deep UV OPA module, 35-40 femto-second pulse duration, tuning range 189-20 μm, pumped with Spectra Physics Spitfire Ace at 5kHz with 1 mJ/pulse will be procured by THz Spectroscopy group. However, the time resolution may still not be sufficient to probe few femto-second dynamics of hot $E_3$ exciton in PbS.